\documentclass[a4paper,fleqn,usenatbib]{mnras}

\usepackage{newtxtext,newtxmath}

\usepackage[T1]{fontenc}
\usepackage{ae,aecompl}

\usepackage{graphicx}	
\usepackage{amsmath}	

\defcitealias{Mathis:1977aa}{MRN}

\title[Dust coagulation in SMGs]{Submillimetre galaxies as laboratories for dust grain
coagulation}

\author[H. Hirashita and C. C. Chen]{
Hiroyuki Hirashita$^{1,2}$\thanks{E-mail: hirashita@asiaa.sinica.edu.tw}
and Chian-Chou Chen$^{1}$\thanks{E-mail: ccchen@asiaa.sinica.edu.tw}
\\
$^{1}$Institute of Astronomy and Astrophysics, Academia Sinica,
Astronomy-Mathematics Building, No.\ 1, Section 4,
Roosevelt Road, Taipei 10617, Taiwan\\
$^{2}$Theoretical Astrophysics, Department of Earth and Space Science, Osaka University,
1-1 Machikaneyama, Toyonaka, Osaka 560-0043, Japan
}

\date{Accepted XXX. Received YYY; in original form ZZZ}

\pubyear{2023}

\begin{document}
\label{firstpage}
\pagerange{\pageref{firstpage}--\pageref{lastpage}}
\maketitle

\begin{abstract}
Coagulation in the dense interstellar medium (ISM) is an important process
that determines the size of the largest grains. We use submillimetre galaxies
(SMGs) as laboratories of grain coagulation, since {some of them}
host the densest ISM on a galactic scale among various populations
of galaxies known. We examine how large the grains can be in such dense environments
based on the mean ISM density estimated from the observed typical dust mass density in SMGs.
We also consider local density enhancement based on a model of supersonic turbulence,
which is expected from strong stellar feedback. In the unlimited coagulation model, in which we
do not impose any coagulation threshold velocity,
grains as large as $\sim 30~\micron$
can form under the
observationally estimated mean gas density
if the Mach number of turbulence is $\mathcal{M}\gtrsim 3$.
We exclude this possibility since the observed emissivity index $\beta\simeq 2$ in the far infrared
(FIR) indicates that such large grains cannot actively form in SMGs.
This means that coagulation does not proceed in an unlimited way:
30-$\micron$ grains should have velocities larger than the coagulation threshold.
If we use a coagulation threshold (upper limit) grain velocity
($\sim 0.08$ km s$^{-1}$) taken from a theoretical study,
grains likely grow only up to $\micron$ size, which is small enough not to
affect the FIR emissivity index.
The above results indicate that SMGs can be used
to constrain the physical processes relevant to coagulation.
\end{abstract}




\section{Introduction}

Dust extinction and emission in galaxies are fundamental processes that alter
the spectral energy distribution (SED) of interstellar radiation field.
In particular, the wavelength dependence of dust extinction is described
by the extinction curve \citep[e.g.][]{Draine:2003ac}, which
is governed by the grain size distributions (distribution functions of
grain radii) as well as
the grain compositions \citep[e.g.][hereafter MRN]{Mathis:1977aa}.
The dust emission SED in galaxies is also affected by the grain size distribution
since stochastically heated small grains, whose temperature distribution function strongly
depends on the grain size, are responsible for the emission
at short wavelengths, especially in the mid-infrared \citep[e.g.][]{Draine:2001aa,Li:2001aa}.
In the far infrared (FIR), the mass absorption coefficient (or emissivity) is not
sensitive to the grain size distribution as long as the grain radii are much smaller than the
wavelengths \citep[e.g.][]{Draine:1984aa}.

Grain growth by coagulation could produce grains large enough to affect the
FIR--submillimetre (submm)
emissivity in dense environments such as protoplanetary discs
\citep[e.g.][]{Testi:2014aa} and protostellar envelopes \citep[e.g.][]{Miotello:2014aa,Wong:2016aa}.
The FIR--submm mass absorption coefficient, $\kappa_\nu$, is well approximated by a power-law
with index $\beta$ as $\kappa_\nu\propto\nu^\beta$ \citep[e.g.][]{Hildebrand:1983aa}.
As the grain size approaches the wavelength, the slope of FIR SED on the
Rayleigh-Jeans side flattens, so that $\beta$ can become $\lesssim 1$ \citep[e.g.][]{Ricci:2010aa}.
If the grains, in contrast, have sizes much smaller than the wavelength,
$\beta$ approaches a value ($\sim 2$) independent of the grain radius
and determined by the grain composition \citep[e.g.][]{Inoue:2020aa}.

In usual galactic interstellar
environments, formation of large grains whose sizes exceed 1 $\micron$ occurs
only locally in molecular cloud cores \citep{Ormel:2009aa,Hirashita:2013aa} as
indicated by observations of coreshine \citep{Pagani:2010aa,Steinacker:2010aa}.
Flat optical--near-infrared extinction curves in circum-nuclear regions in active galactic
nuclei also indicate that such a dense environment may host dust growth
\citep{Maiolino:2001aa,Maiolino:2001ab,Gaskell:2004aa}.
For the formation of larger grains that could affect the FIR emissivity, dense
protostellar environments
are necessary as mentioned above.
These large grains, however, does not influence the
grain size distribution of the entire ISM, since they are readily shattered in the diffuse ISM
\citep{Jones:1994aa,Jones:1996aa,Hirashita:2009ab}. In Milky Way-like galaxies,
the maximum grain radius is determined by the balance between shattering and coagulation
\citep{Huang:2021aa,Chang:2022aa}, and it is consistent with the maximum
radius ($a\sim 0.25~\micron$) as indicated by the extinction curve \citepalias{Mathis:1977aa}.
However, our knowledge of
grain properties is biased to nearby `normal' galaxies.
Since there are distant galaxies whose ISM is in a much denser condition as seen in starbursts,
there is a possibility of galaxy-wide active formation of large grains by coagulation.
Efficient grain growth by coagulation on a galactic scale may change the global dust
properties in galaxies; thus, it could affect dust mass
estimates based on FIR observations.

Among the known galaxy populations, the densest galaxy-scale environments
{could} be seen in {some}
submm galaxies (SMGs). SMGs have kpc-scale extremely high star formation rate
surface densities \citep[e.g.][]{Ikarashi:2015aa},
which reflect exceedingly high gas surface density \citep[e.g.][]{Bouche:2007aa}.
Therefore, it is expected that {some} SMGs host the most `friendly'
environment for dust growth by
coagulation. {The high density is supported in Section \ref{subsec:density}
using actually observed quantities; thus, regardless of how SMGs achieve high density
\citep[see e.g.][and references therein for the
physical mechanisms of creating high dust density environments]{Lovell:2021aa}
the results for coagulation in this paper hold at least for typical SMGs.
We note that our physical properties adopted in this paper are broadly reproduced by
a cosmological simulation \citep{Aoyama:2019aa}.}

The physical properties of SMGs are
often derived from observations of dust emission with assuming dust properties,
since stellar emission is heavily obscured.
Therefore, it is important to clarify the dust size range achieved in SMGs. In particular,
the maximum grain radius could affect the infrared dust properties.

Recent investigations of SMG samples using SED modeling have found FIR
emissivity indices that are similar to those in nearby galaxies
(i.e.\ $\beta\simeq 2$; \citealt{daCunha:2021aa,Cooper:2022aa,Bendo:2023aa,Liao:2023aa}).
Compared with nearby ultraluminous infrared galaxies (ULIRGs)
as used by \citet{Clements:2018aa},
the distribution of SMGs biased to high redshift has an advantage of tracing
short rest-frame wavelengths near to the SED peak with ALMA,
which is capable of resolving the dust distribution scale length important for estimating the
dust mass density (Section \ref{subsec:density}).
Similar values of $\beta$ are obtained by \citet{daCunha:2021aa} for their SMG sample
\citep[see also][]{Casey:2021aa,Cooper:2022aa,Bendo:2023aa}.
Since $\beta$ is practically determined at
$\lambda\gtrsim 200~\micron$ in the SED fitting, the obtained value of $\beta$ means
that the major part of the
grains are unlikely to be larger than $\sim\lambda /(2\upi )\sim 30~\micron$ based on
the Mie theory \citep{Bohren:1983aa}.
Indeed, according to \citep[][figs. 3--5]{Draine:2006aa}, $\beta$ at $\lambda\sim 200~\micron$
starts to deviate from the small grain limit if the maximum grain radius is
10--100~$\micron$ in their power-law grain size distribution.
In fact, the deviation would be clearer in our case, where the grain radius is more concentrated
at a single value (Section \ref{subsec:gsd}).
Therefore, we adopt $a\sim 30~\micron$ for the grain radius above which the emissivity
index should deviate from the small-grain limit ($\beta\simeq 2$), as was also expected
above from the Mie theory.
Since we do not observe $\beta$ deviating from the small-grain limit ($\beta\simeq 2$)
for SMGs in reality,
we are able to clarify rejected cases for the physical conditions relevant for coagulation
(e.g.\ gas density) by examining the formation of 30 $\micron$-sized
grains.

The coagulation rate may be further enhanced because of compressive (supersonic)
turbulence. Some high-redshift star-forming galaxies host dense cold gas with
supersonic velocity dispersion, which is interpreted as turbulent motion
\citep{Swinbank:2011aa,Swinbank:2015aa}.
\citet{Mattsson:2020aa,Mattsson:2020ab} proposed a raised efficiency of dust growth
through the accretion of gas-phase metals
because of the local density enhancement induced by supersonic turbulence.
The increase of accretion efficiency by supersonic turbulence is also shown by
numerical simulations \citep{Li:2020ab}.
Coagulation is also shown to be enhanced in the ISM with supersonic turbulence
\citep{Li:2021aa}. Thus, supersonic turbulence expected to be caused by stellar feedback
in SMGs would potentially enhance large-grain formation by coagulation.

The goal of this paper is to examine how large the grains can be in SMGs.
This also enables us to constrain the physical quantities that govern the coagulation rate, such as
grain velocity, turbulence velocity, and gas density.
{We concentrate on coagulation and neglect other possible grain growth mechanisms.
In particular, dust growth by the accretion of gas-phase metals is neglected for the following two
reasons \citep{Hirashita:2012aa}:
(i) accretion predominantly affects the smallest grains because smaller grains
have larger surface-to-volume ratios. (ii) Accretion stops when all the gas-phase
metals condense on the grain surface, while coagulation continues.}

{In principle, our results in this paper are also applicable to nuclear starburst regions in
some nearby galaxies. We do not consider nearby galaxies in this paper
for the following two reasons: (i) We cannot take advantage of redshift; that is,
as we will argue below, observations that could resolve the relevant region at
rest-frame $\sim 200~\micron$
are crucial, but this wavelength is not accessible by ALMA if redshift is much smaller
than 1. (ii) Compact central regions may
be optically thick at submm wavelengths \citep[e.g.][]{Sakamoto:2021aa}. Thus,
if we concentrate only on the central regions,
it is difficult to obtain the intrinsic spectral index of dust emission.}

This paper is organized as follows.
In Section~\ref{sec:model}, we review the model for coagulation,
and explain how the model is applied to SMGs.
In Section~\ref{sec:result}, we show the results including the dependence on
some principal parameters, and develop analytic formulae. These formulae are used
for further discussion in Section \ref{sec:discussion}.
Finally we give conclusions in Section \ref{sec:conclusion}.

\section{Model}\label{sec:model}

We present the coagulation model that is used to calculate the evolution of grain size distribution.
In particular, we examine how large the grains can be in the dense environments in SMGs.
The coagulation model is based on \citet{Hirashita:2019aa}, but neglecting
the other processes than coagulation. Since the background gas density is important for
coagulation, we also model the gas density including its inhomogeneity.

\subsection{Background density}\label{subsec:density}

The mean background density is estimated from the observed dust mass density in
SMGs. We also consider inhomogeneity in density (or local density enhancement)
induced by supersonic turbulence, which is expected from stellar feedback
\citep[e.g.][]{Hopkins:2011aa}.
The local density enhancement
is expressed by using the mean and mean square hydrogen number densities,
denoted as $\langle n_\mathrm{H}\rangle$ and
$\langle n_\mathrm{H}^2\rangle$, respectively.
We are interested in the density structures induced on a galactic scale, not in
individual star-forming clouds, because we investigate a possibility of
galaxy-wide enhancement of coagulation efficiency in this paper (see the Introduction).

We estimate the mean hydrogen number density from observed quantities.
Assuming that the dust is distributed in a sphere of radius $R_\mathrm{dust}$,
we obtain the mean dust mass density
$\langle\rho_\mathrm{dust}\rangle =M_\mathrm{dust}/(\frac{4}{3}\upi R_\mathrm{dust}^3)$.
Using the dust-to-gas ratio, $\mathcal{D}$, we estimate the mean gas mass density
$\langle\rho_\mathrm{gas}\rangle =\langle\rho_\mathrm{dust}\rangle /\mathcal{D}$, which
is further related to the mean hydrogen number density as
$\langle\rho_\mathrm{gas}\rangle=\mu m_\mathrm{H}\langle n_\mathrm{H}\rangle$,
where $m_\mathrm{H}$ is the mass of hydrogen atom and
the factor $\mu$ accounts for the contribution from helium (we adopt
$\mu =1.4$ throughout this paper). We finally obtain
the mean hydrogen number density as
\begin{align}
\langle n_\mathrm{H}\rangle
=7.0\times 10^2\left(\frac{M_\mathrm{dust}}{10^9~\mathrm{M}_{\sun}}\right)
\left(\frac{R_\mathrm{dust}}{1~\mathrm{kpc}}\right)^{-3}
\left(\frac{\mathcal{D}}{0.01}\right)^{-1}~\mathrm{cm}^{-3}.
\end{align}
Here we adopt $R_\mathrm{dust}\simeq 1$ kpc
\citep{Simpson:2015aa,Ikarashi:2015aa,Hodge:2016aa,Chen:2017aa,Gullberg:2019aa},
and $M_\mathrm{dust}\sim 10^9$ M$_{\sun}$ \citep{daCunha:2015aa,Dudzeviciute:2020aa}.
In this paper, considering the variety in $M_\mathrm{dust}$ and $R_\mathrm{dust}$ among SMGs,
we investigate a wide range of $\langle n_\mathrm{H}\rangle$ (Section \ref{subsec:gsd}).

We assume that the density inhomogeneity is caused by supersonic turbulence, which is
induced by stellar feedback \citep[e.g.][]{Joung:2006aa,Walker:2014aa}.
Supersonic turbulence predicts the density enhancement as a function of
mach number \citep[e.g.][]{Federrath:2008aa,Federrath:2010aa}.
Assuming a lognormal
distribution for the density \citep[e.g.][]{Vazquez:1994aa,Passot:1998aa},
the mean density enhancement is estimated as
\begin{align}
\frac{\langle n_\mathrm{H}^2\rangle}{\langle n_\mathrm{H}^2\rangle^2}=1+b^2\mathcal{M}^2,
\label{eq:enhancement}
\end{align}
where $b\simeq 0.5$ \citep[e.g.][]{Padoan:1997aa} and $\mathcal{M}$ is the mean
Mach number of the turbulence.
Since the rate of grain--grain collisions per unit volume is proportional to
the square of density, the coagulation rate is raised by a factor of
$(1+b^2\mathcal{M}^2)$.

\subsection{Evolution of grain size distribution}\label{subsec:review}

We calculate the evolution of grain size distribution using the coagulation equation
(or the Smoluchowski equation).
The grain size distribution is denoted as $n(a,\, t)$, where $a$ is the grain radius and
$t$ is the time, and is defined such that
$n(a,\, t)\,\mathrm{d}a$ is the number density
of dust grains with radii between $a$ and $a+\mathrm{d}a$.
In the equation, we use the grain mass distribution
$\rho_\mathrm{d}(m,\, t)$, which is defined as
\begin{align}
\rho_\mathrm{d}(m,\, t)\,\mathrm{d}m=\frac{4}{3}\upi a^3sn(a,\, t)\,\mathrm{d}a,
\label{eq:rho_n}
\end{align}
where $m=(4\upi /3)a^3s$ is the grain mass ($s$ is the grain material density).
From this equation, we see that the grain mass contained in a logarithmic bin ($\mathrm{d}\ln a$)
is proportional to $a^4n(a,\, t)$.

We consider the evolution of the spatially averaged grain mass distribution function,
$\langle\rho_\mathrm{d}(m,\, t)\rangle$
(or equivalently $\langle n(a,\, t)\rangle$), taking into account the local
density enhancement described by equation~(\ref{eq:enhancement}).
We assume that the density field of dust traces that of gas on a scale comparable to
the size of the maximum turbulence eddy, which is much smaller than the galaxy scale.
As we will show later, this assumption of grain--gas coupling broadly holds for the fiducial gas density estimated
above (Section \ref{subsec:condition}).
As derived in Appendix \ref{app:coag}, the coagulation equation
\citep[e.g.][]{Hirashita:2019aa} in our problem becomes
\begin{align}
\lefteqn{\frac{1}{1+b^2\mathcal{M}^2}\frac{\partial\langle\rho_\mathrm{d}(m,\, t)\rangle}{\partial t}}
\nonumber\\
&= -m\langle\rho_\mathrm{d}(m,\, t)\rangle\int_0^\infty\alpha (m_1,\, m)
\langle\rho_\mathrm{d}(m_1,\, t)\rangle
\mathrm{d}m_1\nonumber\\
&+ \int_0^\infty\int_0^\infty\alpha (m_1,\, m_2)\langle\rho_\mathrm{d}(m_1,\, t)\rangle
\langle\rho_\mathrm{d}(m_2,\, t)\rangle\nonumber\\
&\times
m_1\delta (m-m_1-m_2)\mathrm{d}m_1\mathrm{d}m_2,\label{eq:coag_mean}
\end{align}
where $\alpha$ is the kernel function, which is expressed as
$\alpha (m_1,\, m_2)=\upi (a_1+a_2)^2v_{1,2}/(m_1m_2)$
using the relative velocity $v_{1,2}$ in the collision between
two grains with masses $m_1$ and $m_2$ (radii $a_1$ and $a_2$,
respectively).
The relative velocity is evaluated in Section \ref{subsec:vel}.
The above equation indicates that coagulation proceeds more quickly by a factor
of $(1+b^2\mathcal{M}^2)$ because of the local density enhancement.

Since the mean dust mass density 
is the integration of $\langle\rho_\mathrm{d}(m,\, t)\rangle$
for $m$, we estimate the dust-to-gas ratio, $\mathcal{D}$, as
\begin{align}
\mathcal{D}=\frac{1}{\langle\rho_\mathrm{gas}\rangle}
\int_0^\infty\langle\rho_\mathrm{d}(m,\, t)\rangle\,\mathrm{d}m.
\label{eq:norm}
\end{align}
In particular, this equation is used for the normalization of the grain size distribution in the
initial condition. Note that coagulation conserves the total dust mass.

\subsection{Modelling the grain velocities}\label{subsec:vel}

We determine the grain velocity as a function of grain radius, denoted as
$v_\mathrm{gr}(a)$,
by the following procedures (appendix C of \citealt{Hirashita:2019aa} and
section 2 of \citealt{Hirashita:2021ab}).
We assume a Kolmogorov spectrum for the turbulence \citep[e.g.][]{Kobayashi:2022aa} with
characteristic eddy size $L_0$ and velocity $v_0$.
Under this assumption, the turnover time of an eddy with velocity $v$ is written as
$\tau_\mathrm{turn}=L_0v^2/v_0^3$.
We estimate the gas drag time-scale for a grain as a function of grain radius
$\tau_\mathrm{dr}(a)$ under a given density $n_\mathrm{H}$ and temperature $T_\mathrm{gas}$.
Assuming that the grain motion is subsonic,\footnote{This is correct for most of the grain radius
range in our paper. Even if grain motion is supersonic for large $a$ and $\mathcal{M}$,
we apply the formula for subsonic case for the purpose of obtaining simple scaling relation.
However, the formulae in this section can be approximately applied
even if the grain velocity is moderately supersonic.}
the drag time-scale (or the grain--gas coupling time-scale), denoted as
$\tau_\mathrm{dr}$, is written as $\tau_\mathrm{dr}=sa/(c_\mathrm{s}\rho_\mathrm{gas})$,
where $c_\mathrm{s}=(k_\mathrm{B}T_\mathrm{gas}/\tilde{\mu})^{1/2}$
is the isothermal sound speed with
$k_\mathrm{B}$ being the Boltzmann constant and $\tilde{\mu}$ being the mean particle weight.
We adopt $\tilde{\mu}=2.2m_\mathrm{H}$ assuming that the hydrogen gas to be fully molecular
(though the results are not significantly altered even if we assume that the gas is fully atomic).
We determine the grain velocity by $\tau_\mathrm{dr}(a)=\tau_\mathrm{turn}$, which means that
a grain effectively interacts with the smallest eddy that can transfer momentum
through the drag force within a turnover time.

Here we choose the values of $L_0$ and $v_0$ suitable for our problem.
We write $v_0=\mathcal{M}c_\mathrm{s}$.
Since the system responds on the dynamical time-scale,
$t_\mathrm{dyn}=1/(G\rho_0)^{1/2}$ ($\rho_0$ is the mass density),
we assume that the turbulence eddy has a size of $L_0=v_0t_\mathrm{dyn}$.
This means that we set $L_0$ to the effective Jeans length, which is also used for
the turbulence model (with $\mathcal{M}=1$) in \citet{Ormel:2009aa}.
We write $\rho_0=q\rho_\mathrm{gas}$, where $q\geq 1$ accounts for the contribution from
components other than gas (such as stars and dark matter).
Effectively, $L_0$ can be taken as the maximum scale for energy input, so that the
turbulence spectrum assumed above is only applicable to eddy sizes $\leq L_0$.

Using the above setup for the turbulence, we solve $\tau_\mathrm{dr}(a)=\tau_\mathrm{turn}$
to obtain the grain velocity as
\begin{align}
v_\mathrm{gr}(a) &=G^{1/4}q^{1/4}
\mathcal{M}s^{1/2}c_\mathrm{s}^{1/2}\rho_\mathrm{gas}^{-1/4}a^{1/2}
\label{eq:vel_org}\\
&= 0.32q^{1/4}\left(\frac{\mathcal{M}}{3}\right)\left(\frac{s}{3.5~\mathrm{g~cm}^{-3}}\right)^{1/2}
\left(\frac{T_\mathrm{gas}}{100~\mathrm{K}}\right)^{1/4}
\nonumber\\
&\times\left(\frac{n_\mathrm{H}}{10^3~\mathrm{cm}^{-3}}\right)^{-1/4}
\left(\frac{a}{1~\micron}\right)^{1/2}~\mathrm{km~s}^{-1}.
\label{eq:vel}
\end{align}
We estimate the relative velocity $v_{1,2}$ by assuming a random direction
in each collision \citep{Hirashita:2013aa}.
We adopt the following approximate estimate for $n_\mathrm{H}$.
Since our model does not really give local $n_\mathrm{H}$, we simply
use the root-mean-square density:
$n_\mathrm{H}=(\langle n_\mathrm{H}^2\rangle )^{1/2}
=(1+b^2\mathcal{M}^2)^{1/2}\langle n_\mathrm{H}\rangle$
(equation \ref{eq:enhancement}), or equivalently
$\rho_\mathrm{gas}=(1+b^2\mathcal{M}^2)^{1/2}\langle\rho_\mathrm{gas}\rangle$.
In reality, the dynamical coupling between dust and gas depends on the detailed small-scale
density structures of the ISM \citep{Hopkins:2016aa}.
This effect is left for future work since a numerical method that treats hydrodynamic evolution of the
ISM is needed. We also neglect grain charges and magnetic fields, since their effects on grain velocities
are minor for large grains ($a\gtrsim 0.1~\micron$), on which we are aiming at putting constraints
\citep{Lee:2017aa}.
In particular, the high velocity induced by turbulence easily overcome the Coulomb barrier
\citep{Hirashita:2019aa}.

In reality, the turbulence has the maximum eddy size ($\sim L_0$) as mentioned above.
For this reason, the grains cannot be accelerated beyond $v_0$, the velocity of the maximum-sized
eddy. If the grain velocity
calculated by equation (\ref{eq:vel}) is larger than $v_0=\mathcal{M}c_\mathrm{s}$,
the grain cannot be coupled with the gas motion on scales smaller than $L_0$.
In other words, this grain is decoupled from the turbulent motion.
Therefore, the condition for grain--gas decoupling is written as
$v_\mathrm{gr}>\mathcal{M}c_\mathrm{s}$ (note that $v_\mathrm{gr}$ is formally estimated
from equation \ref{eq:vel} and that it does not reflect the real grain velocity achieved),
which is translated to
\begin{align}
\langle\rho_\mathrm{gas}\rangle
<\frac{Gqa^2s^2}{c_\mathrm{s}^{2}\,\sqrt{1+b^2\mathcal{M}^2}}
\equiv\langle\rho_\mathrm{gas}\rangle_\mathrm{dec},\label{eq:decoupling}
\end{align}
where $\langle\rho_\mathrm{gas}\rangle_\mathrm{dec}$
is the gas density below which the grain is decoupled from the turbulence. This density is
equivalent to the hydrogen number density, $\langle n_\mathrm{H}\rangle_\mathrm{dec}\equiv
\langle\rho_\mathrm{gas}\rangle_\mathrm{dec}/(\mu m_\mathrm{H}$).

In order to clarify or constrain the physical processes concerned with coagulation,
we take the following approach. First, we adopt the \textit{unlimited coagulation model},
in which we (i) use equation (\ref{eq:vel})
regardless of whether $v_\mathrm{gr}>\mathcal{M}c_\mathrm{s}$ or not
(or equivalently
whether grains and turbulence are coupled or decoupled) and
(ii) assume that coagulation occurs at any collisional velocity.
Both of the assumptions (i) and (ii) are optimistic for grain coagulation.
The second assumption is particularly critical since,
in reality, there is a threshold velocity above which grains cannot stick
(i.e.\ grains are bounced or shattered). Therefore, the unlimited coagulation model
serves to examine the maximum possible grain radius.
The grain radius achieved in the unlimited coagulation model is tested against
the observed absence of 30-$\micron$ grains. This step enables us to
discuss what limits the grain growth in reality. Two physical processes mentioned above,
that is, grain--gas decoupling and coagulation threshold velocity, are examined as limiting factors
for coagulation.
In particular, we aim at providing
a constraint on coagulation threshold velocity purely from SMGs,
which is independent of theoretical \citep{Chokshi:1993aa,Dominik:1997aa,Wada:2013aa}
and experimental \citep{Blum:2000aa} constraints. 

\subsection{Initial condition and parameters}\label{subsec:param}

For the initial condition of grain size distribution, we adopt a
power-law form as suggested by \citetalias{Mathis:1977aa};
that is, $n\propto a^{-3.5}$, which is appropriate for the Milky Way.
The range of the grain radii is chosen to be $a=0.001$--0.25 $\micron$.
As shown by \citet{Hirashita:2019aa}, the grain size distribution tends to converge to the
\citetalias{Mathis:1977aa} shape in metal-enriched environments
such as in the Milky Way.
Some metal-line observations suggest that high-redshift intense starbursts are hosted by
metal-enriched (solar or supersolar) environments
\citep{Rigopoulou:2018aa,Shapley:2020aa,Litke:2022aa}.

Since we do not know precisely the physical condition
realized before the current starburst episode in SMGs, we adopt the
\citetalias{Mathis:1977aa} grain size distribution as one of the
most typical distribution functions for the initial condition.
However, the results below are insensitive to the initial condition after grains
grow to a few $\micron$.
For the dust-to-gas ratio, we adopt $\mathcal{D}=0.01$, which is similar to the value in the
Milky Way \citep[e.g][]{Weingartner:2001aa}.
The total dust-to-gas ratio is used for the normalization of grain
size distribution at $t=0$ (equation \ref{eq:norm}).

The coagulation equation (equation \ref{eq:coag_mean}) is discretized
following appendix B of \citet{Hirashita:2019aa} and the grain radii are considered
in the range of 0.01--1000 $\micron$. We use 128 logarithmically spaced grain radius bins
and adopt $n=0$ (or $\rho_\mathrm{d}=0$) at the maximum and minimum grain radii
for the boundary condition.

We basically fix the following parameters other than $\mathcal{M}$ and
$\langle n_\mathrm{H}\rangle$.
We adopt $s=3.5$ g cm$^{-2}$, which is appropriate
for silicate \citep{Weingartner:2001aa}.
Carbonaceous dust has a lower material density.
Since the number density of dust grains is lower for higher $s$ with a fixed $\mathcal{D}$,
adopting high
material density gives more conservative estimate for coagulation.
We also assume $q=1$, which gives a conservative estimate for the grain velocity, and
adopt $T_\mathrm{gas}=100$~K as a typical gas temperature of the cold ISM
\citep[e.g.][]{Tielens:2005aa}.
The dependences on $q$ and $T_\mathrm{gas}$ are weak with a power of 1/4
(equation~\ref{eq:vel});
thus, compared with the gas density and the Mach number,
these quantities only have a minor influence on the results.
In the end, we concentrate on $\mathcal{M}$ and $\langle n_\mathrm{H}\rangle$.
We consider $\langle n_\mathrm{H}\rangle =700$~cm$^{-3}$ for
the fiducial value according to the estimate in Section~\ref{subsec:density}.
For the density enhancement, we adopt $b=0.5$ as quoted in Section \ref{subsec:density}.
The Mach number is a free parameter, and is purely constrained from the absence of
grains with $a\gtrsim 30~\micron$.

We consider typical ages up to 100--300~Myr based on the duration of the current starburst
\citep{Tacconi:2008aa,daCunha:2015aa,Dudzeviciute:2020aa}.
This time-scale is consistent with that of the buildup of the stellar mass
($10^{11}$~M$_{\sun}$) with a typical star formation rate of
$\sim 10^3$ M$_{\sun}$ yr$^{-1}$ \citep[e.g.][]{Wiklind:2014aa}, and is
comparable to the gas depletion time-scale $\sim 100$~Myr
\citep{Dudzeviciute:2020aa}. Because of these values, we adopt the typical
duration of coagulation as $t=100$ Myr, and consider the evolution of grain size distribution
up to $t=300$ Myr.

\section{Results}\label{sec:result}

\subsection{Evolution of grain size distribution}\label{subsec:gsd}

\begin{figure}
 \includegraphics[width=\columnwidth]{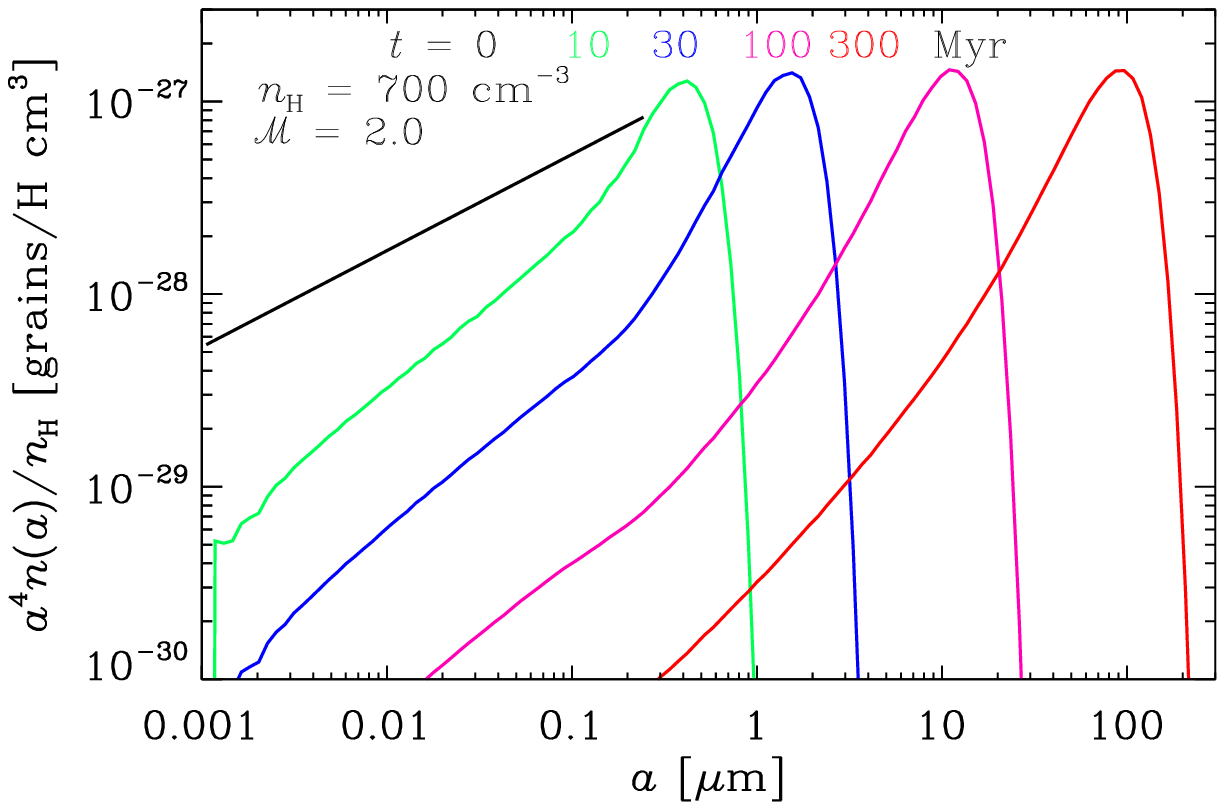}
 \caption{Evolution of grain size distribution for $\mathcal{M}=2$ and
 $\langle n_\mathrm{H}\rangle =700$~cm$^{-3}$. The grain size distribution
 is multiplied by $a^4/n_\mathrm{H}$ so that the resulting quantity is
 proportional to the dust mass density in logarithmic bins per gas mass.
 The grain size distributions are shown for ages
 $t=0$, 10, 30, 100, and 300 Myr, and each age corresponds to the
 colour shown in the legend.}
 \label{fig:gsd}
\end{figure}

We show the evolution of grain size distribution in Fig.~\ref{fig:gsd}
for $\mathcal{M}=2$ and $\langle n_\mathrm{H}\rangle =700$~cm$^{-3}$
at $t=0$--300~Myr. We choose the fiducial value for the mean density, and a moderately
supersonic case $\mathcal{M}=2$ as a tentative value.
As expected, the peak of the
grain size distribution (expressed as `grain mass distribution' on logarithmic scales by multiplying $a^4$;
Section \ref{subsec:review})
shifts to larger radii with time. With $\mathcal{M}=2$,
the grain radius reaches $a\simeq 10~(100)~\micron$ at $t=100$ (300) Myr.
Thus, if coagulation proceeds continuously with an unlimited manner
(as assumed in the unlimited coagulation model; Section~\ref{subsec:vel}),
the grain radii become much larger than those
usually seen in Milky Way-like environments ($\sim 0.25~\micron$; \citetalias{Mathis:1977aa}).

\begin{figure}
 \includegraphics[width=\columnwidth]{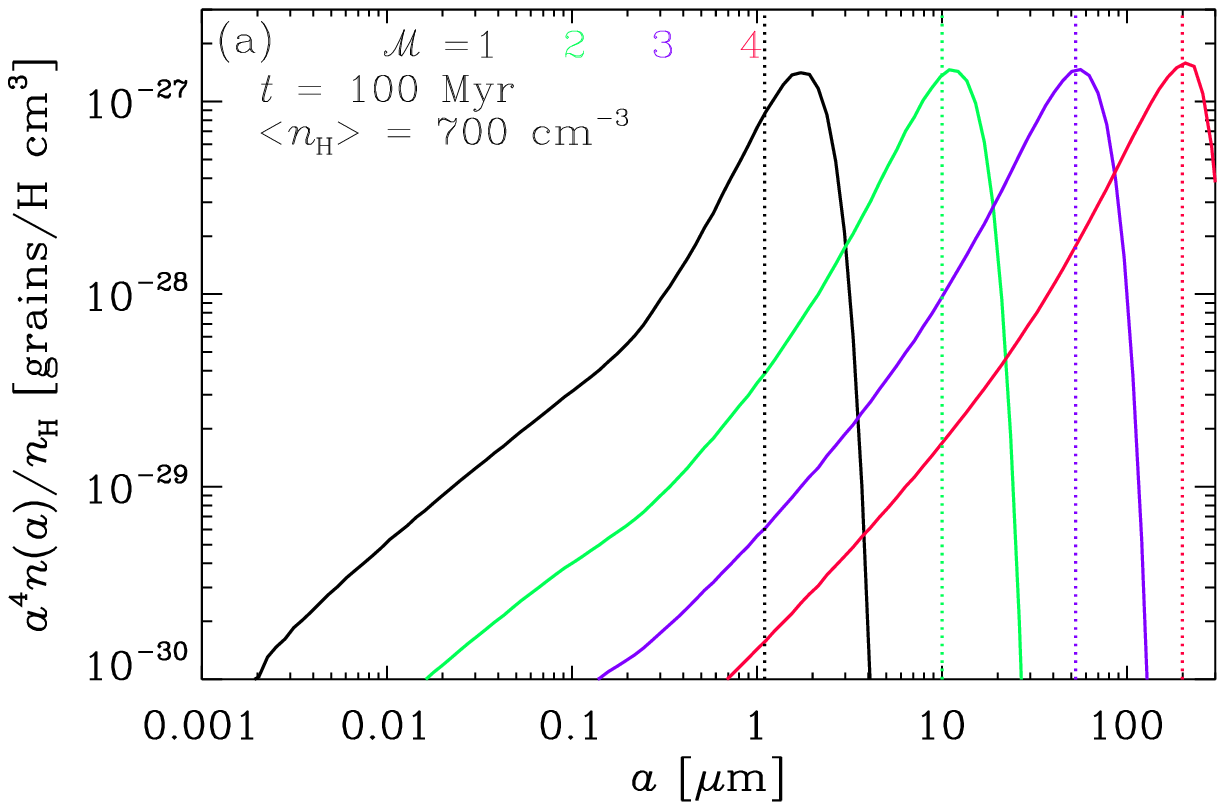}
 \includegraphics[width=\columnwidth]{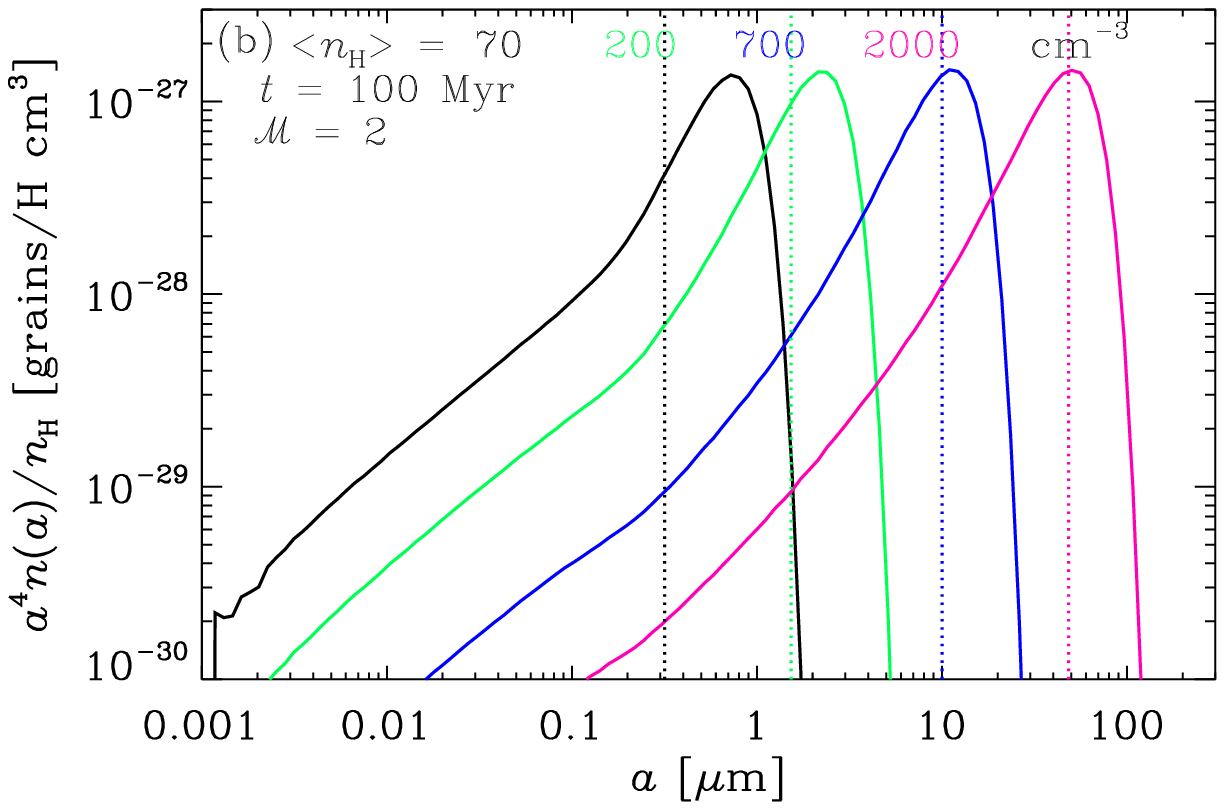}
 \caption{
 Grain size distributions at $t=100$~Myr
 for (a) various $\mathcal{M}$ with $\langle n_\mathrm{H}\rangle =700$~cm$^{-3}$
 and (b) various $\langle n_\mathrm{H}\rangle$ with $\mathcal{M}=2$.
 The grain radii shift towards larger values with increasing $\mathcal{M}$ and
 $\langle n_\mathrm{H}\rangle$, and the correspondence between the values of
 $\mathcal{M}$ and $\langle n_\mathrm{H}\rangle$ and the colours is shown in the legend
 of each panel. In both panels, the vertical dotted lines with the same colours mark
 the grain radii predicted from our analytic formula (equation~\ref{eq:a_ch}) with $\eta =1.5$.
 Note that the analytic formula can be applied at grain radii much larger than 1 $\micron$,
 where the effect of the initial condition is insignificant.}
 \label{fig:gsd_100Myr}
\end{figure}

We vary $\mathcal{M}$ and  $\langle n_\mathrm{H}\rangle$, and compare the results
for various cases at $t=100$ Myr. We examine two cases: $\mathcal{M}=1$--4 with the fiducial density
($\langle n_\mathrm{H}\rangle =700$ cm$^{-3}$), and
$\langle n_\mathrm{H}\rangle =70$--2000 cm$^{-3}$ with a fixed Mach number
($\mathcal{M}=2$). The upper bounds for the values of $\mathcal{M}$ and
$\langle n_\mathrm{H}\rangle$ are large enough to discuss the cases for grain growth
beyond $\sim 30~\micron$.
The resulting grain size distributions are shown in Fig.\ \ref{fig:gsd_100Myr}.

We observe in Fig.\ \ref{fig:gsd_100Myr}a that, as the Mach number increases,
the grain radii shift towards larger values. This is due to the increase of
grain velocity (equation \ref{eq:vel}) and the enhancement of density
(equation \ref{eq:enhancement}).
The mean number density also greatly affects the grain size distribution as
shown in Fig.\ \ref{fig:gsd_100Myr}b. The increase of the mean density has a similar
effect to that of the density enhancement above.

Fig.\ \ref{fig:gsd_100Myr} also shows that
grains larger than 30 $\micron$ can form in the unlimited coagulation model
if $\mathcal{M}\gtrsim 3$ (under the fiducial density)
or if $\langle n_\mathrm{H}\rangle\gtrsim 2000$ cm$^{-3}$ (under $\mathcal{M}=2$).
Thus, to be consistent with the lack of grains with $a\gtrsim 30~\micron$,
$\mathcal{M}\lesssim 3$ and
$\langle n_\mathrm{H}\rangle\lesssim 2000$ cm$^{-3}$ should be satisfied.
As we will discuss in Section \ref{sec:discussion}, however,
it is more likely that grain velocities are so high that coagulation is stopped at
smaller radii.

\subsection{Analytic formulae for the grain radius}\label{subsec:analytic}

The above results indicate that the grain size distribution (more precisely
the grain mass distribution $a^4n$) has a well defined peak after coagulation.
Thus, this peak is defined as the characteristic grain radius, denoted as $a_\mathrm{ch}$.
We derive an analytic formula that gives $a_\mathrm{ch}$.

The evolution of grain size distribution is predominantly affected by
the grain velocity and the gas density, both of which regulate the grain--grain
collision rate. The grain--grain collision rate, $\tau_\mathrm{coll}$, is
estimated as
$\tau_\mathrm{coll}=1/[(1+b^2\mathcal{M}^2)\langle n_\mathrm{gr}\rangle
\upi a_\mathrm{ch}^2v_\mathrm{gr}(a_\mathrm{ch})]$,
where $\langle n_\mathrm{gr}\rangle$ is the mean number density of the grains.
Note that the collisional time-scale is shortened by a factor of $(1+b^2\mathcal{M}^2)$
as shown in Section \ref{subsec:review}.
The mean number density of grains is approximately estimated as
$\langle n_\mathrm{gr}\rangle\simeq \mathcal{D}\langle\rho_\mathrm{gas}\rangle /(\frac{4}{3}
\upi a_\mathrm{ch}^3s)$.
Using equation (\ref{eq:vel_org}),
we obtain the following estimate for the collision time-scale:
\begin{align}
\tau_\mathrm{coll}=
\frac{4s^{1/2}a_\mathrm{ch}^{1/2}}
{3G^{1/4}q^{1/4}(1+b^2\mathcal{M}^2)^{7/8}\mathcal{DM}\langle\rho_\mathrm{gas}\rangle^{3/4}c_\mathrm{s}^{1/2}}.
\end{align}
We assume that the time-scale of grain growth by coagulation ($\tau_\mathrm{coag}$) is regulated by
$\tau_\mathrm{coll}$ as $\tau_\mathrm{coag}=\eta\tau_\mathrm{coll}$ with
$\eta$ being a parameter of order unity.

We expect that at a given time $t$, the grains grow up to a point of
$\tau_\mathrm{coag}\simeq t$. Thus, the characteristic grain radius at age $t$ is
estimated as
\begin{align}
a_\mathrm{ch}\simeq
\frac{9G^{1/2}q^{1/2}(1+b^2\mathcal{M}^2)^{7/4}\mathcal{M}^2\mathcal{D}^2\langle\rho_\mathrm{gas}\rangle^{3/2}c_\mathrm{s}t^2}{16\eta^2s}.\label{eq:a_ch}
\end{align}
Extracting the dependence on $t$, $\mathcal{M}$ and $\langle n_\mathrm{H}\rangle$
with the others fixed to the standard values (Section \ref{subsec:param}), we obtain
$a_\mathrm{ch}=(1.7/\eta^2)(1+b^2\mathcal{M}^2)^{7/4}\mathcal{M}^2
(\langle n_\mathrm{H}\rangle /700~\mathrm{cm}^{-3})^{3/2}(t/100~\mathrm{Myr})^2~\micron$.
We also mark $a_\mathrm{ch}$ predicted from the
above formula in Fig.~\ref{fig:gsd_100Myr}.
We adopt $\eta =1.5$ since this value successfully fits the peak of
$a^4n$ if the grains grow much beyond $a_\mathrm{ch}\sim 1~\micron$.
We note that, if the grains do not grow beyond $\sim 1~\micron$, the effect of the initial condition
is still significant. We observe from Fig.\ \ref{fig:gsd_100Myr} that the above analytic formula
is successful in explaining the peak grain radii as long as they are well above $1~\micron$.
Thus, we use the above analytic formula with $\eta =1.5$ for the following analysis
since we are interested in the formation of large grains in this paper.

\section{Discussion}\label{sec:discussion}

\subsection{Condition for 30-$\micron$ grain formation}\label{subsec:condition}

We examine the condition in which grains grow up to $a\sim 30~\micron$.
In reality, such large grains cannot be dominant in actually observed SMGs; thus, this step
clarifies cases that should be rejected.

\begin{figure}
 \includegraphics[width=\columnwidth]{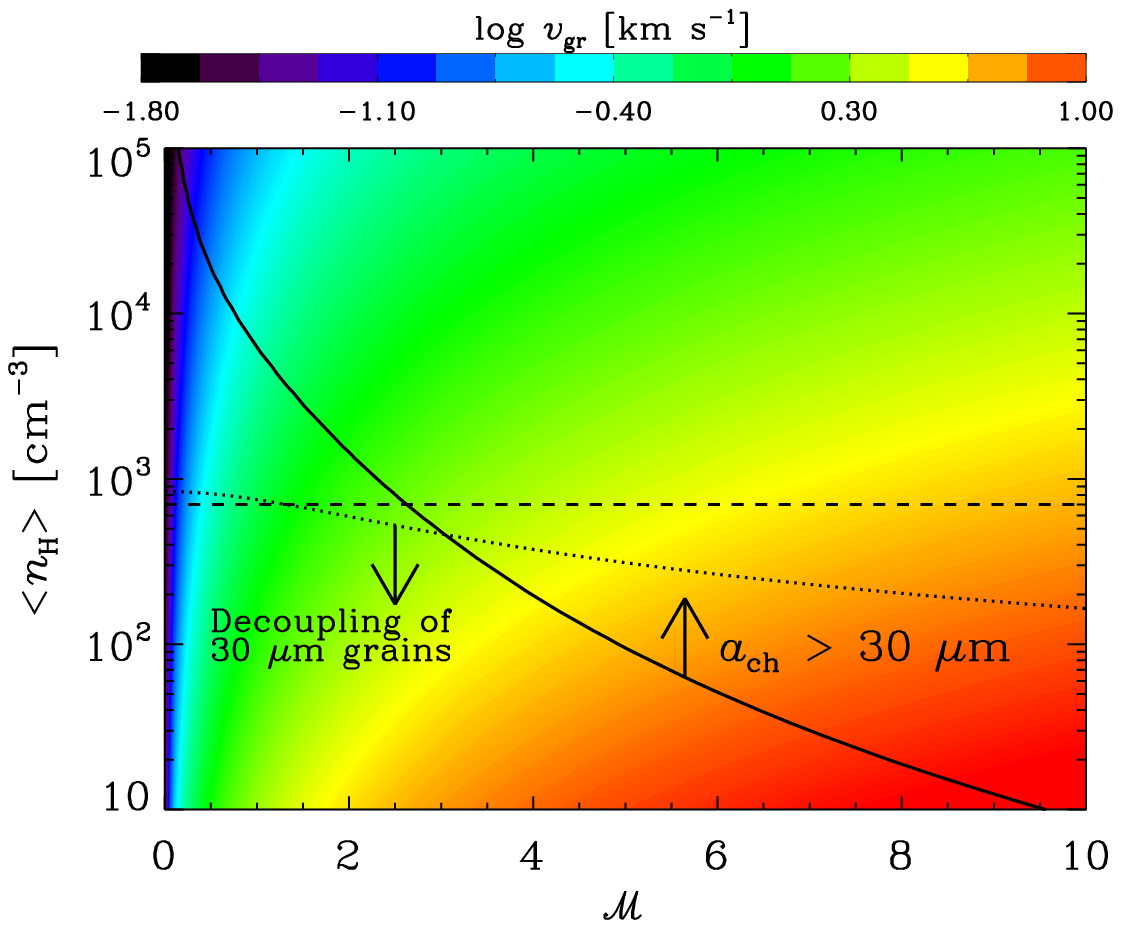}
 \caption{
 Condition for 30-$\micron$ grain formation and associated grain velocities at $t=100$~Myr on
 the mean density $\langle n_\mathrm{H}\rangle$ vs.\ Mach number $\mathcal{M}$ plane.
 We adopt the fiducial values for the
 other parameters. The solid line shows the boundary, above which formation of 30-$\micron$ grains
 is possible in the unlimited coagulation model.
 The dotted line shows the boundary below which 30-$\micron$ grains are decoupled from the
 turbulence. The horizontal dashed line marks the fiducial gas density
 ($\langle n_\mathrm{H}\rangle =700$~cm$^{-3}$). The background colours present the velocity
 predicted from our model (equation \ref{eq:vel}) as shown by the colour bar.
 We observe that the formation of
 30-$\micron$ grains is mostly associated with 1 km s$^{-1}$-level velocities
 (especially if the turbulence is supersonic).
 Thus, the absence
 of such large grains in SMGs indicate either that the parameters ($\mathcal{M}$ and
 $\langle n_\mathrm{H}\rangle$) are below the solid line or that coagulation is
 prohibited at velocities $\gtrsim 1$ km s$^{-1}$. We also observe that the dust grains
 are mostly coupled with the turbulence at and above the fiducial value of
 $\langle n_\mathrm{H}\rangle$.
 }
 \label{fig:cond30}
\end{figure}

In Fig.\ \ref{fig:cond30}, we show the condition for the formation of 30-$\micron$ grains.
We observe that, if both $\langle n_\mathrm{H}\rangle$ and $\mathcal{M}$ are large enough,
grains can grow beyond $a=30~\micron$ in the unlimited coagulation model.
Thus, the observed absence of 30-$\micron$ grains can be due to a Mach number
or a density much lower than the fiducial value. The latter possibility is not likely since
we used observed quantities to estimate the gas (dust) density. We also find that,
at the fiducial gas density with supersonic turbulence,
the grains are coupled with the turbulence even if they are as large
as $a\sim 30~\micron$. Thus, decoupling is not likely to be the reason why such large grains
are not observed in SMGs.

We also present the grain velocity as a function of
$\langle n_\mathrm{H}\rangle$ and $\mathcal{M}$ in the background colours
of Fig.\ \ref{fig:cond30}.
We find that the formation of 30-$\micron$ grains is associated with grain
velocities $\gtrsim 1$ km s$^{-1}$ for most of the parameter range
(neglecting the very high $\langle n_\mathrm{H}\rangle\gtrsim 10^4$ cm$^{-3}$
and subsonic regime, which is far from the fiducial density but is still discussed
in Section \ref{subsec:suggested_radii}).
Thus, if the real ISM in SMGs is in the regime
above the line of $a_\mathrm{ch}=30~\micron$ (solid line),
the observed absence of such large grains
means that the grains do not coagulate at such large velocities.
Indeed, with a km s$^{-1}$-level velocity, grains could be shattered
\citep{Tielens:1994aa,Jones:1994aa,Jones:1996aa}.
This velocity is much higher than the sticking threshold velocity, typically
$\sim 1$ m s$^{-1}$, given by \citet{Chokshi:1993aa} and
\citet{Dominik:1997aa}, but experimental results show that grains still stick
at higher velocities \citep{Blum:2000aa}.
With numerical experiments, \citet{Wada:2013aa} showed that grains composed
of 0.1-$\micron$-sized monomers start to be disrupted if the collision velocities
exceed 80 m s$^{-1}$. Thus, it is reasonable to presume that coagulation is prohibited
at km s$^{-1}$ velocities.

Another possible reason for the absence of a 30-$\micron$ grain population is simply
that the parameters $\langle n_\mathrm{H}\rangle$ and $\mathcal{M}$ are below the
line of $a_\mathrm{ch}=30~\micron$ (solid line) in Fig.\ \ref{fig:cond30}.
This means that the turbulence is subsonic or moderately supersonic
($\mathcal{M}\lesssim 3$ at $\langle n_\mathrm{H}\rangle =700$ cm$^{-3}$).
However, since 30-$\micron$ grains have large velocities in most of the parameter space,
it is more likely that grain velocity larger than the coagulation threshold is the primary reason for
the absence of such an extremely large grain population.

The above results may also indicate that, if $\mathcal{M}$ is large, grain velocities
can be high enough for shattering to occur. In this case, small grains should form.
In particular, polycyclic aromatic hydrocarbons (PAHs) form through shattering of
large carbonaceous grains \citep{Hirashita:2020aa,Narayanan:2023aa}.
Another fragmentation mechanisms such as rotational disruption
could take place in environments with high radiation
field \citep[e.g.][]{Hoang:2019aa}.
Therefore, observations of PAH emission in the mid-infrared will clarify if
shattering, which counteracts coagulation, is actively occurring in SMGs.

\subsection{Suggested grain radii from a coagulation threshold velocity}\label{subsec:suggested_radii}

In the above, we have assumed unlimited coagulation to maximize the possibility of extremely
large grains. Here, we set a coagulation threshold velocity above which grains do not
coagulate in collision. We use the analytic formulae developed in Section \ref{subsec:analytic}.

A simple way to examine the effect of the coagulation threshold velocity is to solve
$v_\mathrm{gr}(a)=v_\mathrm{th}$, where $v_\mathrm{th}$ is the coagulation threshold velocity.
This equation is solved for $a$
with given values of $\langle n_\mathrm{H}\rangle$ and $\mathcal{M}$ (and with
the other parameters fixed to the fiducial values).
The solution of $a$
gives an estimate for the maximum grain radius that could be formed by coagulation.
This radius is denoted as $a_\mathrm{max}$. On the other hand, even with unlimited coagulation,
grains can grow up to $a_\mathrm{ch}$ at most. Thus, we expect that the radius grains achieve
can be estimated as $a=\min (a_\mathrm{max},\, a_\mathrm{ch})$.

As discussed in Section \ref{subsec:condition}, there is a variety in experimentally
and theoretically obtained
coagulation threshold velocity. We adopt a relatively large value for the coagulation
threshold velocity based on \citet{Wada:2013aa}, who simulated collisions between
large grains composed of 0.1 $\micron$-sized monomers. Since they showed that
collisions with relative velocities $\gtrsim 0.08$ km s$^{-1}$ start to disrupt grains,
we adopt $v_\mathrm{th}=0.08$ km s$^{-1}$. If we adopt further lower coagulation threshold
velocities as shown in experimental studies by \citet{Blum:2000aa}, coagulation is stopped
at grain radii smaller than 0.1 $\micron$, which cannot explain the flattening of extinction
curves in dense regions \citep{Hirashita:2014ab}. Thus, we adopt the relatively high
coagulation threshold, which is still lower than our constraint obtained for SMGs
($\sim 1$ km s$^{-1}$; Section \ref{subsec:condition}).

\begin{figure}
 \includegraphics[width=\columnwidth]{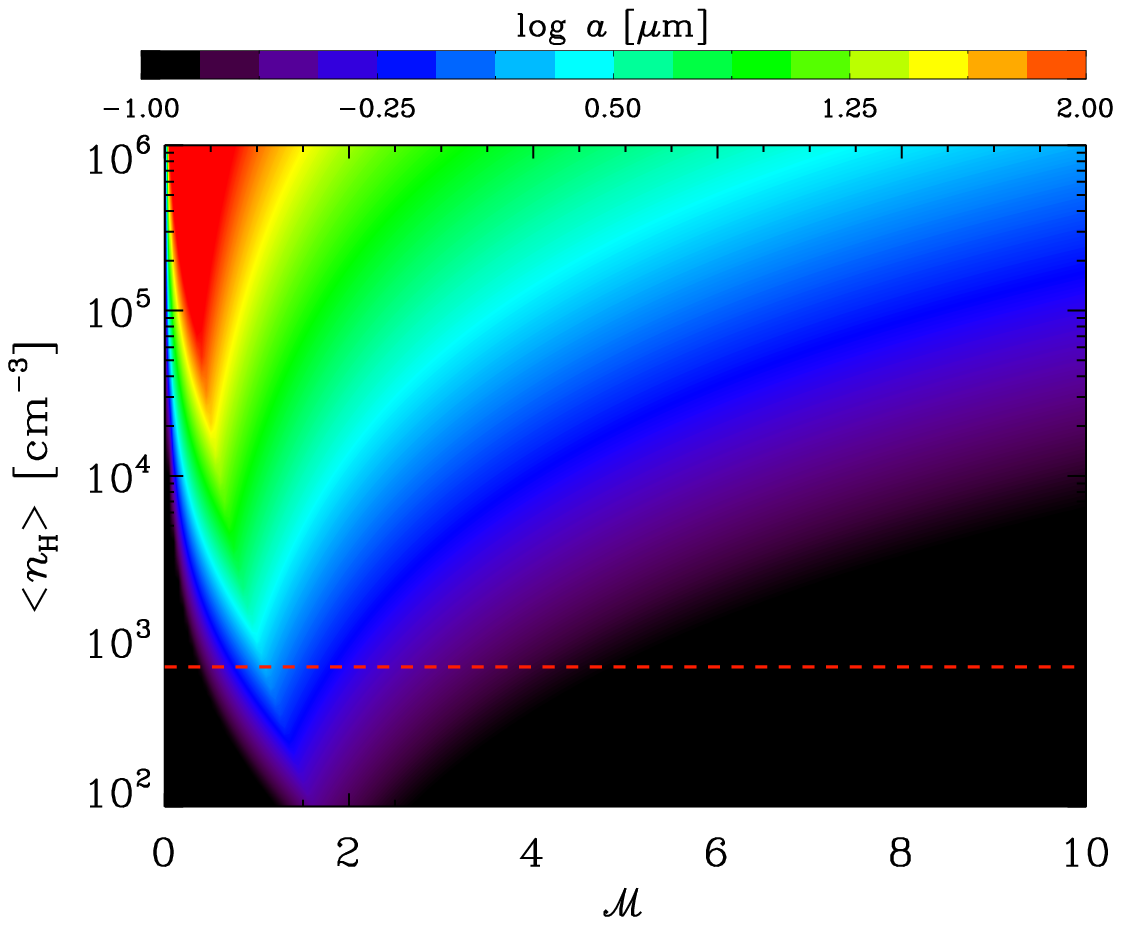}
 \caption{Grain radius expected to be achieved in 100 Myr for various values of
 $\langle n_\mathrm{H}\rangle$ and $\mathcal{M}$ with a coagulation threshold
 velocity of $v_\mathrm{th}=0.08$ km s$^{-1}$. We adopt the values specified
 in Section \ref{subsec:param} for the
 other parameters. The expected grain radius,
 $a=\min (a_\mathrm{max},\, a_\mathrm{ch})$, is colour-coded as shown by
 the colour bar. Note that, if the estimated $a$ is smaller than the initial grain radii
 (typically 0.1 $\micron$ in the \citetalias{Mathis:1977aa} grain size distribution), the effect of
 coagulation is negligible. Thus, we only show the area with $a\gtrsim 0.1~\micron$ in colour
 (i.e.\ the black region shows the area where coagulation has little impact).}
 \label{fig:radii_achieved}
\end{figure}

In Fig.\ \ref{fig:radii_achieved}, we show $a=\min (a_\mathrm{max},\, a_\mathrm{ch})$.
We observe that, at $\langle n_\mathrm{H}\rangle =700$ cm$^{-3}$,
grains can grow up to a few $\micron$ if $\mathcal{M}\sim 1$--2.
At smaller and larger $\mathcal{M}$, the turbulence velocity is, respectively, too low
(i.e.\ inefficient in grain growth) and too high (i.e.\ collisions do not cause coagulation),
so that the grain radii are smaller. As the density becomes higher,
the grain radii tend to be larger.
We also find a regime where grains can grow
up to a few tens of microns with subsonic turbulence at
$\langle n_\mathrm{H}\rangle\gtrsim\text{a few}\times 10^4$ cm$^{-3}$.
The absence of 30-$\micron$-sized grains
in SMGs indicate that such dense and quiescent regions cannot be dominant in
hosting the total dust mass.

\section{Conclusions}\label{sec:conclusion}

We investigate a possibility of using SMGs as laboratories for dust grain coagulation,
since {some} SMGs host the densest galaxy-scale ISM among the known galaxy populations.
Observations indicate that a major part of the dust grains should not have
radii larger than 30 $\micron$. We calculate the coagulation equation to clarify
the grain radius that could be achieved in appropriate conditions for the ISM in SMGs.

Before theoretical calculations, we estimate the mean density of the ISM in SMGs
from the observed galaxy radius and dust mass under an assumption that
SMGs have dust-to-gas ratios similar to the Milky Way (or a solar-metallicity environment).
Based on this estimate, we adopt the fiducial mean density as
$\langle n_\mathrm{H}\rangle =700$ cm$^{-3}$.
We further consider density inhomogeneity caused by supersonic turbulence,
which is expected
from stellar feedback. The grain velocities driven by turbulence is modelled
under given values of density and Mach number.
We do not include any velocity limit for coagulation, so that our model, which
is referred to as the unlimited coagulation model, could investigate
the maximum dust growth that could potentially happen in the given ISM condition.

We find for the unlimited coagulation model
that, in a typical duration of the SMG phase ($\sim 100$ Myr),
grains grow up to $a\sim 30~\micron$  under the fiducial density
($\langle n_\mathrm{H}\rangle =700$ cm$^{-1}$) if $\mathcal{M}\gtrsim 3$.
Alternatively, such large grains could form under lower
$\mathcal{M}\sim 2$ if the density is higher than
$\langle n_\mathrm{H}\rangle\gtrsim 2\times 10^3$ cm$^{-3}$.
However, we also find that this growth is associated with grain velocities
$\sim 1$ km s$^{-1}$, which is too high for grains to coagulate.
This supports the view that coagulation stops before the grains grow up to
30 $\micron$, and explains the absence of 30-$\micron$ grains in SMGs.
In particular, experimental and theoretical studies suggest that coagulation is prohibited
if grain velocities are so large.
Our `astronomical' constraint using SMGs is independent of the experimental and
theoretical constraints, and we
demonstrate that galaxy-scale observations can be used to put a constraint on
coagulation processes.

We also examine the expected grain radius in a more realistic case where
the coagulation threshold velocity is
chosen to be 0.08 km s$^{-1}$. We find that
coagulation can form $\micron$-sized grains at the fiducial density if
$\mathcal{M}\sim 1$--2. With lower and higher $\mathcal{M}$,
grains are smaller because of less frequent grain--grain collision
and because of too high grain velocities for coagulation, respectively.
We also find that in a regime of very high mean density
$\langle n_\mathrm{H}\rangle\gtrsim\text{a few}\times 10^4$ cm$^{-3}$ with subsonic turbulence,
grains can grow up to tens of microns if such a region is sustained for 100 Myr.
This implies that such dense and subsonic regions are not dominant in the ISM of SMGs.

Note that the above results are applied to a dust population dominant in a galaxy, and
do not exclude the existence of extremely large grains in localized dense gas
(e.g.\ star-forming gas). 
More precisely, we are interested in the dust population that
affects the galaxy-scale emissivity index, based on which the possibility of coagulation
beyond 30 $\micron$ is excluded.
At the same time, our models and analytic formulae developed in this paper is
applicable to any galaxy-scale dense enrivonments for the purpose of using them as laboratories of
coagulation.

\section*{Acknowledgements}

{We are grateful to the anonymous referee for useful comments.}
HH thanks the National Science and Technology Council of Taiwan (NSTC) 
for support through grant
111-2112-M-001-038-MY3,
and the Academia Sinica for Investigator Award AS-IA-109-M02.
C.-C.C. acknowledges support from the NSTC through grants
NSTC 109-2112-M-001-016-MY3 and 111-2112M-001-045-MY3, as well as Academia Sinica
through the Career Development Award (AS-CDA-112-M02).

\section*{Data availability}
The data underlying this article will be shared on reasonable request to the corresponding author.



\bibliographystyle{mnras}
\bibliography{/Users/hirashita/bibdata/hirashita}


\appendix

\section{Coagulation equation with local density enhancement}\label{app:coag}

We explain the derivation of equation (\ref{eq:coag_mean}).
The evolution of $\rho_\mathrm{d}$ by coagulation at
position $\bmath{x}$ and time $t$ is described by
\citep[e.g.][]{Hirashita:2019aa}
\begin{align}
\frac{\upartial\rho_\mathrm{d}(m,\,\bmath{x},\, t)}{\upartial t}
= -m\rho_\mathrm{d}(m,\, \bmath{x},\, t)\int_0^\infty\alpha (m_1,\, m)
\rho_\mathrm{d}(m_1,\,\bmath{x},\, t)
\mathrm{d}m_1\nonumber\\
+ \int_0^\infty\int_0^\infty\alpha (m_1,\, m_2)\rho_\mathrm{d}(m_1,\,\bmath{x},\,t)
\rho_\mathrm{d}(m_2,\,\bmath{x},\, t)
m_1\delta (m-m_1-m_2)\mathrm{d}m_1\mathrm{d}m_2.\label{eq:coag}
\end{align}
Here we neglect the advection of dust in space (or equivalently assume that
the advection term cancels out after the following averaging process).
We take the averages for $\bmath{x}$ on both sides of this equation so that we omit
$\bmath{x}$ hereafter.
Under an assumption that the grains are dynamically coupled with the gas,
we obtain for arbitrary grain masses $m'$ and $m''$
\begin{align}
\frac{\langle\rho_\mathrm{d}(m',\, t)\rho_\mathrm{d}(m'',\, t)\rangle}
{\langle\rho_\mathrm{d}(m',\, t)\rangle\langle\rho_\mathrm{d}(m'',\, t)\rangle}=
\frac{\langle n_\mathrm{H}^2\rangle}{\langle n_\mathrm{H}\rangle^2}=
1+b^2\mathcal{M}^2,
\end{align}
where we use equation (\ref{eq:enhancement}) for the second step.
Thus, after applying the averaging procedure for
equation (\ref{eq:coag}), we obtain equation (\ref{eq:coag_mean}).

\bsp	
\label{lastpage}
\end{document}